\documentclass{article}

\usepackage[preprint]{neurips_2023}

\usepackage[utf8]{inputenc} 
\usepackage[T1]{fontenc}    
\usepackage{hyperref}       
\usepackage{url}            
\usepackage{booktabs}       
\usepackage{amsfonts}       
\usepackage{nicefrac}       
\usepackage{microtype}      
\usepackage{xcolor}         
\usepackage{multirow}
\usepackage[inkscapelatex=false]{svg}
\usepackage{natbib}
\setcitestyle{numbers,square}
\usepackage{amsmath}
\usepackage{caption}

\title{MTKD: Multi-Teacher Knowledge Distillation for Image Super-Resolution}

\author{%
  Yuxuan Jiang,
Chen Feng,
Fan Zhang, and 
David Bull \\
Visual Information Laboratory \\
University of Bristol \\
Bristol, BS1 5DD, UK \\
  \texttt{\{yuxuan.jiang, chen.feng, fan.zhang, dave.bull\}@bristol.ac.uk} \\
}

\begin{document}

\maketitle

\begin{abstract}
  Knowledge distillation (KD) has emerged as a promising technique in deep learning, typically employed to enhance a compact student network through learning from their high-performance but more complex teacher variant. When applied in the context of image super-resolution, most KD approaches are modified versions of methods developed for other computer vision tasks, which are based on training strategies with a single teacher and simple loss functions. In this paper, we propose a novel Multi-Teacher Knowledge Distillation (MTKD) framework specifically for image super-resolution. It exploits the advantages of multiple teachers by combining and enhancing the outputs of these teacher models, which then guides the learning process of the compact student network. To achieve more effective learning performance, we have also developed a new wavelet-based loss function for MTKD, which can better optimize the training process by observing differences in both the spatial and frequency domains. We fully evaluate the effectiveness of the proposed method by comparing it to five commonly used KD methods for image super-resolution based on three popular network architectures. The results show that the proposed MTKD method achieves evident improvements in super-resolution performance, up to 0.46dB (based on PSNR), over state-of-the-art KD approaches across different network structures. The source code of MTKD will be made available \url{here} for public evaluation.
\end{abstract}

\section{Introduction}
\label{sec:intro}
  
Image super-resolution (ISR) is an important research topic in image processing; its purpose is to create a high-resolution (HR) image with improved perceptual quality and richer spatial detail from a corresponding low-resolution (LR) version. It is widely used in applications including medical imaging \cite{wang2020deep}, image restoration \cite{liang2021swinir}, enhancement \cite{singh2014various}, and picture coding \cite{afonso2018video}. The past decade has seen impressive performance improvements due to extensive research in this area, in particular associated with advances in deep learning techniques. Learning-based ISR approaches can be classified according to their basic network structure \cite{aleissaee2023transformers}. One major class is based on convolutional neural networks (CNNs), with notable examples including \emph{e.g.}, SRCNN \cite{dong2015image}, EDSR \cite{lim2017enhanced} and RCAN \cite{zhang2018image, lin2022revisiting}. A second class employs Vision Transformer (ViT) networks with important contributions such as SwinIR \cite{liang2021swinir}, Swin2SR \cite{conde2022swin2sr} and HAT \cite{chen2023hat}.

Although these learning-based ISR algorithms have demonstrated superior performance over conventional methods based on classic signal processing theory, they are typically associated with high computational complexity and memory requirements, often inhibiting their practical deployment. To address this issue, research has focused on the development of lightweight ISR methods for real-world applications \cite{jiang2023compressing, cai2022real}. These works often obtain compact models through model compression \cite{buciluǎ2006model} or other simplification \cite{liang2021swinir} of their corresponding full versions. However, when these lite ISR models are directly optimized based on the same training material used for their original counterparts, they generally achieve lower super-resolution performance due to their reduced model capacity and limited learning ability. 

\begin{figure}[t]
  \centering
  \includegraphics[width=1\linewidth]{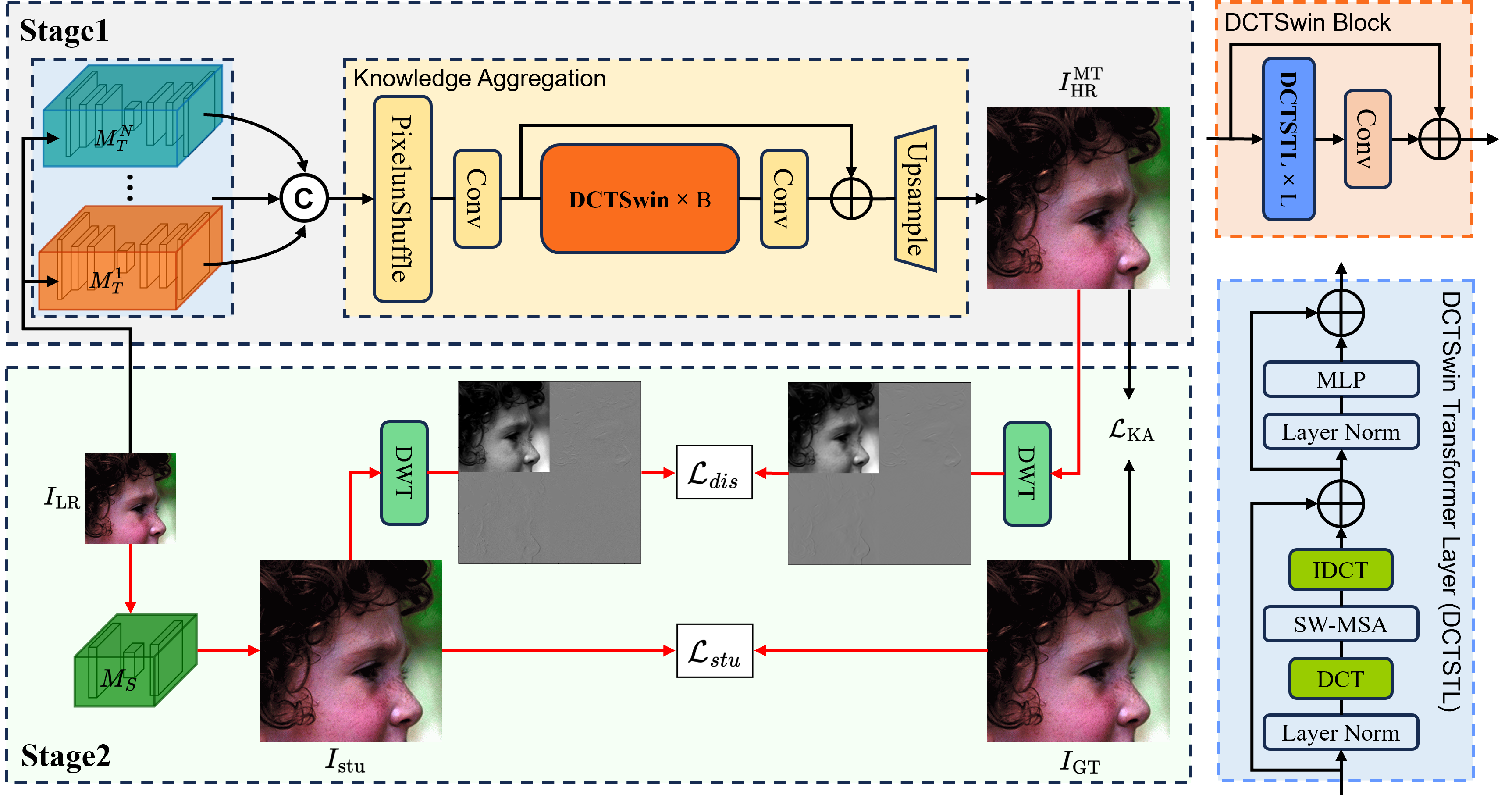}
    \caption{Illustration of the proposed Multi-Teacher Knowledge Distillation framework.}
    \label{fig:multiKDfw}
\end{figure}

To further improve the performance of these low complexity learning-based ISR methods, knowledge distillation (KD) techniques \cite{hinton2015distilling} have been commonly applied. These employ a larger network as a ``teacher'' that transfers knowledge to a smaller ``student'' network. In this process, an auxiliary loss function is often employed to instruct the student to mimic the teacher's output, through minimizing the disparity between intermediate features or by aligning their final predictions. This type of approach has demonstrated promising outcomes for various applications such as image classification \cite{jin2023multi, miles2023understanding_AAAI}, video compression \cite{peng2023accelerating}, object detection \cite{shi2020proxylesskd, zhu2023scalekd}, natural language processing (NLP) \cite{jiao2019tinybert, liu2023pre, kang2023distill} and quality assessment \cite{feng2023rankdvqa}. 

Many prior studies \cite{fang2022cross, zhang2023data, he2020fakd,morris2023st} in KD predominantly focus on the transfer of knowledge from a single teacher to its corresponding student. This strategy loses the advantages of well-performing complex models. They also employ simple loss objectives, such as L1,  to minimize the output difference between the teacher and student models; this does not fully reflect the nature of image super-resolution in reconstructing lost high-frequency information in the higher-resolution content.

In this context, inspired by the advances in multi-teacher selection based KD for natural language processing tasks \cite{yuan2021reinforced} and the classic wavelet transforms \cite{heil1989continuous}, this paper presents a novel Multi-Teacher Knowledge Distillation (MTKD) framework for image super-resolution based on a new wavelet-inspired loss function. As illustrated in Fig. \ref{fig:multiKDfw}, MTKD employs a new Discrete Cosine Transform Swin transformer (DCTSwin) based network to combine the outputs of multiple ISR teacher models and generates an enhanced representation of the high-resolution image, which is then used to guide the student model during distillation. For improving knowledge transfer performance, we have designed a new distillation loss function based on the discrete wavelet transform, which compares the output of the student and the enhanced representation within different frequency subbands. The primary contributions of this work are summarized as follows.

\begin{itemize}
    \item[1)] We, for the first time, propose a novel Multi-Teacher Knowledge Distillation (MTKD) framework for image super-resolution. This framework allows the use of multiple teacher models with different network architectures, which improves the efficiency and diversity of transferred knowledge. 
    
    \item[2)] We have developed a knowledge aggregation network based on novel DCTSwin blocks, which is employed to combine the outputs of multiple teachers to produce a refined representation for ISR knowledge distillation. 

    \item[3)] We have designed an ISR-specific and wavelet-based loss function which collects information from different frequency subbands allowing the model to effectively learn high-frequency information from the original images, which can enhance the performance of knowledge distillation for image super-resolution.
\end{itemize}

We have demonstrated the superior performance of the proposed method through quantitative and qualitative evaluations based on three teacher and student models with distinct network architectures. Our MTKD approach has been compared against five existing KD methods and achieved consistent and evident performance gains, up to 0.46dB assessed by PSNR.

\section{Related Work}
\label{sec:RelatedWorks}

Our work is closely related to two main research topics in the literature: image super-resolution and knowledge distillation.

\subsection{Image Super-Resolution (ISR)}
\label{subsubsec:ISR}

ISR is an image restoration technique, which aims to reconstruct high spatial resolution images from their low-resolution counterparts. In conventional methods, this is achieved using upsampling filters \cite{bull2021intelligent}. However, learning-based ISR methods are predominate due to their superior performance. One of the earliest contributions in this area is SRCNN \cite{dong2015image}, which is based on a simple three-layer convolutional network. This has been further enhanced by incorporating more sophisticated network structures, with notable examples such as VDSR \cite{kim2016accurate} based on residual connections, SRRetNet \cite{ledig2017photo} and EDSR \cite{lim2017enhanced} employing residual blocks, and RCAN \cite{zhang2018image} containing channel weight attention mechanisms. Recently, more effective ISR algorithms have been proposed inspired by the Vision Transformer (ViT) networks \cite{vaswani2017attention}, which leverage self-attention mechanisms for capturing extensive contextual interaction information. Important contributions include the ESRT \cite{lu2022transformer}, SwinIR \cite{liang2021swinir} and Swin2SR \cite{conde2022swin2sr} based on shifted window attention \cite{liu2021swin}, and HAT \cite{chen2023hat} which combines both channel attention and window-based self-attention schemes. To improve the perceptual quality of super-resolution results, generative models such as variational autoencoders (VAEs), generative adversarial networks, and diffusion models have also been exploited, with typical examples including SR-VAE \cite{liu2020photo}, VDVAE-SR \cite{chira2022image}, SRGAN \cite{ledig2017photo}, CAL-GAN \cite{park2023content}, SR3 \cite{saharia2022image} and IDM \cite{gao2023implicit}. For a more comprehensive overview of image super-resolution, readers are referred to references including \cite{yang2017image, lepcha2023image}. 

\subsection{Knowledge Distillation (KD)}
\label{subsubsec:KD}

The objective of Knowledge Distillation (KD) is to improve the model generalization of a compact student model by emulating the behavior of a large well-performing teacher network. It is typically applied together with model compression in order to reduce the complexity of a deep neural network \cite{buciluǎ2006model}. KD offers enhanced performance compared to directly optimizing the compact model on training data only \cite{hinton2015distilling}. Existing KD approaches can be categorized into two primary classes that either: (i)  perform direct emulation of the output from the teacher model (output-level) \cite{jin2023multi, zhang2023data} or (ii) allow the student network to learn intermediate features generated by the teacher model (feature-level) \cite{he2020fakd, fang2022cross, zagoruyko2016paying}. These approaches have been widely applied to many high-level tasks, including, image classification \cite{jin2023multi, miles2023understanding_AAAI, lin2022knowledge, chen2021distilling}, object detection \cite{shi2020proxylesskd, chawla2021data, chen2017learning, zhu2023scalekd} and natural language processing (NLP) \cite{jiao2019tinybert, liu2023pre, kang2023distill}. In the context of ISR, SRKD is a pioneer work \cite{gao2018image} employing KD for image super-resolution. It has been further enhanced through distilling second-order statistical information from feature maps as in FAKD \cite{he2020fakd}, which improves the effective transfer of structural knowledge from the teacher to the student. In \cite{lee2020learning}, the teacher model is replaced by the privileged information obtained from the ground truth high-resolution images, which has been reported to offer improved knowledge distillation performance. Moreover, knowledge distillation has been improved in \cite{zhang2023data} through image zooming and invertible data augmentations, which enhances the generalization of the student model. 

It is noted that all the aforementioned knowledge distillation approaches for ISR are based on a single-teacher framework. In the context of multi-task learning, some recent studies have identified the potential of using multiple teachers for knowledge distillation \cite{gu2021class, meng2021multi, li2020knowledge, jacob2023online}. Similar approaches have also been proposed for applications with single tasks, such as \cite{chebotar2016distilling, wu2019multi, zhu2018knowledge}. These approaches involve leveraging the weighted average of teacher models for distilling into the student model with fixed weights. In contrast, \cite{yuan2021reinforced} proposed an approach that dynamically assigns weights to teacher models based on individual examples. However, these knowledge distillation paradigms have only been investigated for high-level tasks such as classification and detection, and have not been exploited for image super-resolution in the literature.

Moreover, existing knowledge distillation methods typically employ simple loss functions to perform observation in the spatial or feature domain. Due to the nature of the ISR task, which aims to recover spatial details (high-frequency information) in high-resolution images, it is more important to conduct an assessment in the frequency domain during the training process. It is noted that in the literature several token-mixing models have been proposed recently, which replace self-attention modules in transformers with Fourier \cite{lee2105fnet} and wavelet transforms \cite{jeevan2022wavemix}. These models show evident performance enhancement for high-level vision tasks such as image classification and semantic segmentation, while maintaining low computational complexity. In the field of ISR, feature extraction has been performed in the frequency domain in order to achieve competitive performance \cite{zhang2022swinfir, jeevan2024wavemixsr, xu2022dct}, and there are also several attempts to design transform-based (DCT or Fast Fourier Transform) loss functions \cite{lopez2022attention, yadav2021frequency}. However,  as we are aware, wavelet-based loss objectives have not previously been investigated for ISR.

\section{Proposed method: MTKD}
\label{sec:Proposedmethod}

As illustrated in Fig. \ref{fig:multiKDfw}, our Multi-Teacher Knowledge Distillation (MTKD) framework consists of two primary stages: 1) knowledge aggregation and 2) model distillation. In Stage 1, the input low-resolution image $I_\mathrm{LR}\in \mathbb{R}^{H \times W \times C_\mathrm{in}}$ ($H$, $W$ and $C_\mathrm{in}$ represent the image height, width and the number of color channels respectively) is reconstructed by $N$ pre-trained ISR teacher models, denoted by ${M}^1_T \dots M^N_{T}$, resulting in $N$ high-resolution images, $I^{1}_\mathrm{HR} \dots I^{N}_\mathrm{HR}$. These are then concatenated and fed into a Knowledge Aggregation module, which outputs an enhanced high-resolution image, $I^\mathrm{MT}_\mathrm{HR}$. Once the Knowledge Aggregation module is optimized, its parameters are then fixed in Stage 2 when the student model is trained through knowledge distillation.

\subsection{Stage 1: Knowledge Aggregation}
\label{subsec:KnowledgeAggregation}

\subsubsection{Network architecture} The architecture of the network employed for knowledge aggregation is shown in Fig. \ref{fig:multiKDfw}. The input of the network is the concatenation of $N$ teachers' outputs, $[I^{1}_\mathrm{HR} \dots I^{N}_\mathrm{HR}] \in \mathbb{R}^{sH \times sW \times C_\mathrm{in} \times N}$, in which $s$ is the up-scale factor in ISR. A pixel unshuffle layer (together with an additional $3 \times 3$ convolution layer) is then used to down-sample the spatial resolution of the input by a factor of $s$. The down-sampling factor here is the same as the super-resolution up-scale rate in order to keep the following DCTSwin blocks independent from the upscale factor. The feature set output by the convolutional layer is denoted by $F_{s} \in \mathbb{R}^{H \times W \times C}$, where $C$ represents the feature channel number that is a fixed hyperparameter. $F_{s}$ are  then processed by $B$ DCTSwin blocks, generating  deep features $F^{1}_{d},  F^{2}_{d} \dots F^{B}_{d}$. The last set of features $F^{B}_{d}$ is fed into a $3 \times 3$ convolutional layer to produce $F_{d}$, which will be combined with the shallow features $F_{s}$ before being upsampled to the full image resolution. The upsample module is implemented based on the sub-pixel convolution layer \cite{shi2016real}.

\subsubsection{DCTSwin block} As shown in Fig. \ref{fig:multiKDfw}, each DCTSwin block contains $L$ DCTSwin Transformer Layers (DCTSTLs), which are modified from the original Swin Transformer layer \cite{liu2021swin}. It first normalizes the input using a Layer Norm before performing Discrete Cosine Transform (DCT). The transformed coefficients are then processed by the Shifted Window Multi-head Self-Attention (SW-MSA) module \cite{liu2021swin}. The output of the SW-MSA module is further fed into the Inverse DCT (IDCT) module to recover the features in the spatial domain, which are combined with the input of this DCTSTL to achieve residual learning. Here to reduce the number of model parameters and facilitate fast DCT operation, the input of the DCT is segmented into $W_s\times W_s$ blocks prior to the SW-MSA module through window partition \cite{liu2021swin}, and the window reverse operation is performed after the IDCT module to reshape and assemble the output.

\subsubsection{Loss function} To train the Knowledge Aggregation module, we optimize its network parameters by minimizing the L1 loss between the ground-truth image, $I_{GT}$, and the output of the module, $I^\mathrm{MT}_\mathrm{HR}$:
\begin{equation}
  \mathcal{L}_\mathrm{KA} = \left \| I_\mathrm{GT} - I^\mathrm{MT}_\mathrm{HR} \right \|_{1}.
\end{equation}

\subsection{Stage 2: Model Distillation}
\label{subsec:ModelDistillation}

In Stage 2, the network parameters of the optimized Knowledge Aggregation module will be fixed for student-teacher distillation. Here our new MTKD approach does not require the student network to have a similar architecture to one of the teacher models. This allows us to employ teacher models with diverse network architectures in order to achieve improved knowledge distillation performance.

\subsubsection{Loss function} The output of the Knowledge Aggregation module and the ground-truth image are jointly employed to train the student model. Specifically, we first compare the output of the student model $I_\mathrm{stu}$ with the ground truth based on L1 loss:
\begin{equation}
  \mathcal{L}_{stu} = \left \| I_\mathrm{stu} - I_\mathrm{GT} \right \|_{1}.
\end{equation}
The design of the distillation loss is inspired by \cite{zhang2022wavelet}, where a wavelet-based training methodology was employed for image-to-image translation. It is calculated between $I_\mathrm{stu}$ and $I^\mathrm{MT}_\mathrm{HR}$ after decomposing both using a discrete wavelet transform (DWT):
\begin{align}
    \mathcal{L}_{dis} & = \frac{1}{3K+1} \sum_{i,k} \left  \| \mathrm{DWT}_{i,k}(I_\mathrm{stu}) - \mathrm{DWT}_{i,k}(I^\mathrm{MT}_\mathrm{HR}) \right \|_{1},
\end{align}
in which $i\in \{\mathrm{LL, LH, HL, HH}\}$. Here $k=1 \dots K$, which stands for the DWT decomposition level.

These two losses are then combined as the overall loss function $\mathcal{L}_{total}$ as given below.
\begin{align}
    \mathcal{L}_{total} & = \alpha\mathcal{L}_{stu}(I_\mathrm{stu}, I_\mathrm{GT}) + \mathcal{L}_{dis}(I_{stu}, I^\mathrm{MT}_\mathrm{HR}),
\end{align}
where $\alpha$ is a tunable weight to determine the contributions of two losses. It is noted here that we did not employ a wavelet-based loss for $\mathcal{L}_{stu}$. This is because (i) L1 loss is the most commonly used loss function to minimize the difference between the model output and the ground truth; (ii) due to the small $\alpha$ value used, the contribution of $\mathcal{L}_{stu}$ is rather limited. Using L1 loss here can effectively reduce the training complexity without compromising the performance.

\section{Experiment Configuration}
\label{sec:Experiments}

\begin{table}[t]
\small
\centering
\setlength\tabcolsep{6pt}
\caption{The configurations of the employed teacher and student ISR networks.}
\begin{tabular}{r|c|c|c|c}
\toprule
Model & Role & Channel & Block/Group & \#Params(M) \\ \midrule
{SwinIR \cite{liang2021swinir}} & Teacher & 180 & 6/- & 11.8 \\
 {SwinIR\_lightweight \cite{liang2021swinir}}&  Student & 60 & 4/- & 0.9 \\ \midrule
{RCAN \cite{zhang2018image}} & Teacher & 64 & 20/10 & 15.6 \\
{RCAN\_lightweight \cite{fang2022cross, zhang2023data}} & Student & 64 & 6/10 & 5.2 \\ \midrule
{EDSR \cite{lim2017enhanced}} & Teacher & 256 & 32/- & 43 \\
{EDSR\_baseline \cite{lim2017enhanced}} & Student & 64 & 16/- & 1.5 \\ \bottomrule
\end{tabular}
\label{tbl:NetworkCfg}
\end{table}

\subsection{Teacher and student networks}
\label{subsec:Backbone}

To evaluate the effectiveness of the proposed MTKD approach, we followed the evaluation practice in \cite{zhang2023data} by selecting two widely used CNN-based ISR models: EDSR \cite{lim2017enhanced}, RCAN \cite{zhang2018image}, and one popular transformer-based model: SwinIR \cite{liang2021swinir} for comparison. Their open-source full original models have been used here as the teacher networks. As the aim of this paper is to develop and validate a new knowledge distillation method, rather than design a novel lightweight model, we employed the existing low complexity variants of these teacher networks as the compact student models. For EDSR and SwinIR, the student models are the same as in their original papers: EDSR\_baseline  \cite{lim2017enhanced} and SwinIR\_lightweight \cite{liang2021swinir}. For RCAN, due to the lack of the lite model in the corresponding literature, we follow the configurations in \cite{fang2022cross, zhang2023data} to obtain the compact RCAN model, denoted by RCAN\_lightweight. The configuration details of these teacher and student models are summarized in Table \ref{tbl:NetworkCfg}. To fully evaluate the ISR performance, we trained and evaluated each model for super-resolution tasks with multiple scale factors, including $\times 2$, $\times 3$ and $\times 4$.

\subsection{Training configurations}
\label{subsec:Datasets}

In alignment with \cite{liang2021swinir, lim2017enhanced, zhang2018image}, we utilized 800 images from DIV2K \cite{timofte2017ntire} in this experiment for model training. In the training process, we perform random cropping of LR patches with dimensions $64 \times 64$ from the LR images, and the corresponding HR patches are cropped from the ground-truth images based on the scale factor. To achieve data augmentation, random rotation and horizontal flipping are further applied to the training material.

Other training configurations include ADAM optimizer \cite{kingma2014adam} with parameter settings $\beta_1 = 0.9$, $\beta_2 = 0.999$, and $\epsilon = 10^{-8}$; the block number $B$ is 4, layer number $L$ is 2 and the window size $W_s$ is set to 8; the feature channel number $C$ is 24, and the maximum DWT decomposition level $K$ is 1; the factor $\alpha$ used to balance the distillation loss function is set as 0.1. The training batch size is 16 with a total of $2.5 \times 10^5$ iterations; the initial learning rate is set to $10^{-4}$ and is decayed by a factor of 10 at every $10^5$ update. This experiment is implemented on the BasicSR \cite{basicsr} platform using an NVIDIA V100 GPU.

\subsection{Evaluation configurations}

Four commonly used test sets, Set5 \cite{bevilacqua2012low}, Set14 \cite{zeyde2012single}, BSD100 \cite{martin2001database}, and Urban100 \cite{huang2015single}, were employed here for benchmarking the model performance. The ISR performance is assessed using two widely used quality metrics including peak signal-to-noise ratio (PSNR) and structural similarity index (SSIM) \cite{wang2004image}.

To benchmark the performance of the proposed MTKD method, we employed five existing knowledge distillation methods for comparison including basic KD \cite{hinton2015distilling}, AT \cite{zagoruyko2016paying}, FAKD \cite{he2020fakd}, DUKD \cite{zhang2023data}, and CrossKD \cite{fang2022cross}. We also include the results of the corresponding pre-trained teacher and student (training from scratch without any KD) models for benchmarking.

\begin{table}[!ht]
\small
\centering
\caption{Quantitative results (PSNR/SSIM) for the RCAN\_lightweight model.}
\begin{tabular}{c|c|c|c|c|c|c}
\toprule
\multicolumn{2}{c}{} & Dataset & Set5 & Set14 & BSD100 & Urban100 \\ \midrule
Model & Scale & Method & PSNR/SSIM & PSNR/SSIM & PSNR/SSIM & PSNR/SSIM \\ \midrule
\multirow{27}{*}{\rotatebox{270}{RCAN}} & \multirow{9}{*}{$\times$2} & MT & 38.41/0.9626 & 34.49/0.9252 & 32.53/0.9045 & 33.91/0.9429 \\
 &  & Full & 38.27/0.9614 & 34.12/0.9216 & 32.41/0.9027 & 33.34/0.9384 \\
 &  & Compact & 38.07/0.9608 & 33.62/0.9183 & 32.20/0.9000 & 32.32/0.9302 \\ \cmidrule{3-7}
 &  & KD & 38.18/0.9611 & 33.83/0.9197 & 32.29/0.9010 & 32.67/0.9329 \\
 &  & AT & 38.13/0.9610 & 33.70/0.9187 & 32.25/0.9005 & 32.48/0.9313 \\
 &  & FAKD & 38.16/0.9611 & 33.82/0.9190 & 32.27/0.9010 & 32.53/0.9320 \\
 &  & DUKD & \textcolor{blue}{38.23}/\textcolor{blue}{0.9614} & \textcolor{blue}{33.90}/\textcolor{blue}{0.9201} & \textcolor{blue}{32.33}/\textcolor{blue}{0.9016} & \textcolor{blue}{32.87}/\textcolor{blue}{0.9349} \\
 &  & CrossKD & 38.18/0.9612 & 33.82/0.9195 & 32.29/0.9012 & 32.69/0.9331 \\
 &  & \textbf{MTKD (ours)} & \textcolor{red}{38.26}/\textcolor{red}{0.9619} & \textcolor{red}{34.09}/\textcolor{red}{0.9219} & \textcolor{red}{32.40}/\textcolor{red}{0.9031} & \textcolor{red}{33.06}/\textcolor{red}{0.9364} \\ \cmidrule{2-7} \specialrule{0em}{0.1pt}{0.1pt} \cmidrule{2-7}
 & \multirow{9}{*}{$\times$3} & MT & 34.96/0.9315 & 30.93/0.8531 & 29.48/0.8154 & 29.85/0.8822 \\
 &  & Full & 34.74/0.9299 & 30.65/0.8482 & 29.32/0.8111 & 29.09/0.8702 \\
 &  & Compact & 34.56/0.9284 & 30.41/0.8438 & 29.16/0.8076 & 28.48/0.8600 \\  \cmidrule{3-7}
 &  & KD & 34.61/0.9291 & 30.47/0.8447 & 29.21/0.8080 & 28.62/0.8612 \\
 &  & AT & 34.55/0.9287 & 30.43/0.8438 & 29.17/0.8070 & 28.43/0.8577 \\
 &  & FAKD & 34.65/0.9291 & 30.45/0.8442 & 29.21/0.8087 & 28.52/0.8602 \\
 &  & DUKD & \textcolor{blue}{34.74}/\textcolor{blue}{0.9296} & \textcolor{blue}{30.54}/\textcolor{blue}{0.8458} & \textcolor{blue}{29.25}/\textcolor{blue}{0.8088} & \textcolor{blue}{28.79}/\textcolor{blue}{0.8646} \\
 &  & CrossKD & 34.66/0.9291 & 30.50/0.8448 & 29.22/0.8082 & 28.64/0.8617 \\
 &  & \textbf{MTKD (ours)} & \textcolor{red}{34.78}/\textcolor{red}{0.9306} & \textcolor{red}{30.59}/\textcolor{red}{0.8483} & \textcolor{red}{29.34}/\textcolor{red}{0.8106} & \textcolor{red}{29.18}/\textcolor{red}{0.8704} \\ \cmidrule{2-7} \specialrule{0em}{0.1pt}{0.1pt} \cmidrule{2-7}
 & \multirow{9}{*}{$\times$4} & MT & 32.83/0.9027 & 29.06/0.7934 & 27.93/0.7496 & 27.44/0.8232 \\
 &  & Full & 32.63/0.9002 & 28.87/0.7889 & 27.77/0.7436 & 26.82/0.8087 \\
 &  & Compact & 32.32/0.8964 & 28.69/0.7840 & 27.63/0.7381 & 26.34/0.7933 \\  \cmidrule{3-7}
 &  & KD & 32.45/0.8980 & 28.76/0.7860 & 27.67/0.7400 & 26.49/0.7980 \\
 &  & AT & 32.31/0.8967 & 28.69/0.7839 & 27.64/0.7385 & 26.29/0.7927 \\
 &  & FAKD & 32.46/0.8983 & 28.75/0.7859 & 27.68/0.7402 & 26.42/0.7973 \\
 &  & DUKD & \textcolor{blue}{32.56}/\textcolor{blue}{0.8990} & \textcolor{blue}{28.83}/\textcolor{blue}{0.7870} & \textcolor{blue}{27.72}/\textcolor{blue}{0.7410} & \textcolor{blue}{26.62}/\textcolor{blue}{0.8020} \\
 &  & CrossKD & 32.45/0.8984 & 28.81/0.7866 & 27.69/0.7406 & 26.53/0.7992 \\
 &  & \textbf{MTKD (ours)} & \textcolor{red}{32.62}/\textcolor{red}{0.9009} & \textcolor{red}{28.84}/\textcolor{red}{0.7901} & \textcolor{red}{27.88}/\textcolor{red}{0.7447} & \textcolor{red}{27.08}/\textcolor{red}{0.8108}  \\ \bottomrule
\end{tabular}
\label{tbl:RCANResults}
\end{table}

\section{Results and Discussion}
\label{sec:Results}

\begin{table}[!t]
\small
\centering
\caption{Quantitative results (PSNR/SSIM) for the EDSR\_baseline model.}
\begin{tabular}{c|c|c|c|c|c|c}
\toprule
\multicolumn{2}{c}{} & Dataset & Set5 & Set14 & BSD100 & Urban100 \\ \midrule
Model & Scale & Method & PSNR/SSIM & PSNR/SSIM & PSNR/SSIM & PSNR/SSIM \\ \midrule
\multirow{24}{*}{\rotatebox{270}{EDSR}} & \multirow{8}{*}{$\times$2} & MT & 38.41/0.9626 & 34.49/0.9252 & 32.53/0.9045 & 33.91/0.9429 \\
 &  & Full & 38.11/0.9601 & 33.92/0.9195 & 32.32/0.9013 & 32.93/0.9351 \\
 &  & Compact & 37.96/0.9608 & 33.55/0.9176 & 32.17/0.9003 & 31.99/0.9274 \\  \cmidrule{3-7}
 &  & KD & 37.97/\textcolor{blue}{0.9610} & \textcolor{blue}{33.60}/0.9180 & 32.19/0.9002 & 32.09/0.9283 \\
 &  & AT & 37.99/0.9607 & 33.58/0.9173 & 32.21/0.8996 & 32.08/0.9275 \\
 &  & FAKD & \textcolor{blue}{38.01}/0.9604 & 33.59/\textcolor{blue}{0.9181} & \textcolor{blue}{32.23}/\textcolor{blue}{0.9008} & \textcolor{blue}{32.11}/\textcolor{blue}{0.9292} \\
 &  & \textbf{MTKD (ours)} & \textcolor{red}{38.08}/\textcolor{red}{0.9612} & \textcolor{red}{33.82}/\textcolor{red}{0.9196} & \textcolor{red}{32.29}/\textcolor{red}{0.9017} & \textcolor{red}{32.42}/\textcolor{red}{0.9308} \\ \cmidrule{2-7} \specialrule{0em}{0.1pt}{0.1pt} \cmidrule{2-7}
 & \multirow{8}{*}{$\times$3} & MT & 34.96/0.9315 & 30.93/0.8531 & 29.48/0.8154 & 29.85/0.8822 \\
 &  & Full & 34.65/0.9282 & 30.52/0.8462 & 29.25/0.8093 & 28.80/0.8653 \\
 &  & Compact & 34.36/0.9273 & 30.28/0.8421 & 29.09/0.8066 & 28.14/0.8528 \\  \cmidrule{3-7}
 &  & KD & 34.39/\textcolor{blue}{0.9277} & 30.31/0.8427 & 29.10/\textcolor{blue}{0.8071} & 28.19/\textcolor{blue}{0.8533} \\
 &  & AT & 34.40/0.9268 & 30.30/\textcolor{blue}{0.8431} & \textcolor{blue}{29.16}/0.8056 & 28.12/0.8519 \\
 &  & FAKD & \textcolor{blue}{34.47}/0.9273 & \textcolor{blue}{30.37}/0.8425 & 29.12/0.8062 & \textcolor{blue}{28.21}/\textcolor{blue}{0.8533} \\
 &  & \textbf{MTKD (ours)} & \textcolor{red}{34.54}/\textcolor{red}{0.9286} & \textcolor{red}{30.48}/\textcolor{red}{0.8450} & \textcolor{red}{29.20}/\textcolor{red}{0.8086} & \textcolor{red}{28.48}/\textcolor{red}{0.8578} \\ \cmidrule{2-7} \specialrule{0em}{0.1pt}{0.1pt} \cmidrule{2-7}
 & \multirow{8}{*}{$\times$4} & MT & 32.83/0.9027 & 29.06/0.7934 & 27.93/0.7496 & 27.44/0.8232 \\
 &  & Full & 32.46/0.8968 & 28.80/0.7876 & 27.71/0.7420 & 26.64/0.8033 \\
 &  & Compact & 32.09/0.8944 & 28.56/0.7814 & 27.57/0.7372 & 26.03/0.7849 \\  \cmidrule{3-7}
 &  & KD & 32.12/0.8952 & \textcolor{blue}{28.56}/0.7823/ & 27.56/\textcolor{blue}{0.7382} & 26.02/0.7861 \\ 
 &  & AT & 32.08/0.8935 & 28.49/0.7798 & 27.51/0.7365 & 26.00/0.7866 \\
 &  & FAKD & \textcolor{blue}{32.21}/\textcolor{blue}{0.8957} & 28.55/\textcolor{blue}{0.7826} & \textcolor{blue}{27.58}/0.7377 & \textcolor{blue}{26.11}/\textcolor{blue}{0.7892} \\
 &  & \textbf{MTKD (ours)} & \textcolor{red}{32.29}/\textcolor{red}{0.8967} & \textcolor{red}{28.73}/\textcolor{red}{0.7849} & \textcolor{red}{27.69}/\textcolor{red}{0.4707} & \textcolor{red}{26.32}/\textcolor{red}{0.7918} \\ \bottomrule
\end{tabular}
\label{tbl:EDSRResults}
\end{table}

\begin{table}[t]
\small
\centering
\caption{Quantitative results (PSNR/SSIM) for the SwinIR\_lightweight model.}
\begin{tabular}{c|c|c|c|c|c|c}
\toprule
\multicolumn{2}{c}{} & Dataset & Set5 & Set14 & BSD100 & Urban100 \\ \midrule
Model & Scale & Method & PSNR/SSIM & PSNR/SSIM & PSNR/SSIM & PSNR/SSIM \\ \midrule
\multirow{24}{*}{\rotatebox{270}{SwinIR}} & \multirow{8}{*}{$\times$2} & MT & 38.41/0.9626 & 34.49/0.9252 & 32.53/0.9045 & 33.91/0.9429 \\
 &  & Full & 38.35/0.9620 & 34.14/0.9227 & 32.44/0.9030 & 33.40/0.9393 \\
 &  & Compact & 38.14/0.9611 & 33.86/0.9206 & 32.31/0.9012 & 32.76/0.9340 \\  \cmidrule{3-7}
 &  & KD & 38.15/\textcolor{red}{0.9619} & \textcolor{blue}{33.90}/0.9211 & \textcolor{blue}{32.33}/0.9023 & 32.79/0.9342 \\
 &  & FAKD & \textcolor{blue}{38.16}/\textcolor{blue}{0.9615} & 33.87/\textcolor{blue}{0.9216} & 32.34/\textcolor{blue}{0.9024} & \textcolor{blue}{32.81}/\textcolor{blue}{0.9345} \\
 &  & AT & 38.14/\textcolor{blue}{0.9615} & 33.85/0.9209 & 32.32/0.9021 & 32.74/0.9341 \\
 &  & DUKD & 38.13/0.9610 & 33.78/0.9194 & 32.26/0.9007 & 32.63/0.9327 \\
 &  & \textbf{MTKD (ours)} & \textcolor{red}{38.21}/\textcolor{red}{0.9619} & \textcolor{red}{34.03}/\textcolor{red}{0.9218} & \textcolor{red}{32.39}/\textcolor{red}{0.9030} & \textcolor{red}{32.92}/\textcolor{red}{0.9351} \\ \cmidrule{2-7} \specialrule{0em}{0.1pt}{0.1pt} \cmidrule{2-7}
 & \multirow{8}{*}{$\times$3} & MT & 34.96/0.9315 & 30.93/0.8531 & 29.48/0.8154 & 29.85/0.8822 \\
 &  & Full & 34.89/0.9312 & 30.77/0.8503 & 29.37/0.8124 & 29.29/0.8744 \\
 &  & Compact & 34.62/0.9289 & 30.54/0.8463 & 29.20/0.8082 & 28.66/0.8624 \\ \cmidrule{3-7}
 &  & KD & 34.61/0.9292 & 30.55/\textcolor{blue}{0.8469} & 29.23/0.8100 & \textcolor{blue}{28.67}/0.8627 \\
 &  & FAKD & \textcolor{blue}{34.65}/\textcolor{blue}{0.9296} & \textcolor{blue}{30.56}/0.8463 & \textcolor{blue}{29.25}/\textcolor{blue}{0.8113} & 28.65/\textcolor{blue}{0.8629} \\
 &  & AT & 34.59/0.9289 & 30.53/0.8464 & 29.22/0.8094 & 28.63/0.8618 \\
 &  & DUKD & 34.55/0.9285 & 30.53/0.8456 & 29.20/0.8080 & 28.53/0.8604 \\
 &  & \textbf{MTKD (ours)} & \textcolor{red}{34.70}/\textcolor{red}{0.9300} & \textcolor{red}{30.66}/\textcolor{red}{0.8480} & \textcolor{red}{29.31}/\textcolor{red}{0.8116} & \textcolor{red}{29.03}/\textcolor{red}{0.8679} \\ \cmidrule{2-7} \specialrule{0em}{0.1pt}{0.1pt} \cmidrule{2-7}
 & \multirow{8}{*}{$\times$4} & MT & 32.83/0.9027 & 29.06/0.7934 & 27.93/0.7496 & 27.44/0.8232 \\
 & & Full & 32.72/0.9021 & 28.94/0.7914 & 27.83/0.7459 & 27.07/0.8164 \\
 &  & Compact & 32.44/0.8976 & 28.77/0.7858 & 27.69/0.7406 & 26.47/0.7980 \\ \cmidrule{3-7}
 &  & KD & \textcolor{blue}{32.43}/\textcolor{blue}{0.8984} & 28.78/\textcolor{blue}{0.7862} & \textcolor{blue}{27.71}/0.7426 & 26.50/0.7981 \\
 &  & FAKD & \textcolor{blue}{32.43}/0.8979 & \textcolor{blue}{28.79}/0.7857 & 27.69/\textcolor{blue}{0.7427} & \textcolor{blue}{26.51}/\textcolor{blue}{0.7982} \\
 &  & AT & 32.42/0.8982 & 28.77/0.7860 & 27.70/0.7421 & 26.47/0.7981 \\
 &  & DUKD & 32.41/0.8973 & \textcolor{blue}{28.79}/0.7860 & 27.69/0.7405 & 26.43/0.7972 \\
 &  & \textbf{MTKD (ours)} & \textcolor{red}{32.52}/\textcolor{red}{0.8993} & \textcolor{red}{28.87}/\textcolor{red}{0.7885} & \textcolor{red}{27.79}/\textcolor{red}{0.7450} & \textcolor{red}{26.85}/\textcolor{red}{0.8071} \\ \bottomrule
\end{tabular}
\label{tbl:SwinIRResults}
\end{table}

\subsection{Quantitative Evaluation}
\label{subsec:QuantitativeE}

The quantitative results for three compact ISR networks, EDSR\_baseline  \cite{lim2017enhanced}, SwinIR\_lightweight \cite{liang2021swinir} and RCAN\_lightweight \cite{fang2022cross, zhang2023data} that have been trained with different knowledge distillation methods, including basic KD \cite{hinton2015distilling}, AT \cite{zagoruyko2016paying}, FAKD \cite{he2020fakd}, DUKD \cite{zhang2023data}, CrossKD \cite{fang2022cross} and the proposed MTKD are presented in Table \ref{tbl:RCANResults}-\ref{tbl:SwinIRResults}, in which the best and second-best performers are highlighted in red and blue, respectively. Here we also provide results for the original EDSR, SwinIR and RCAN models (denoted by ``full'' in  Table \ref{tbl:RCANResults}-\ref{tbl:SwinIRResults}), their pre-trained student variants (without knowledge distillation, denoted by ``compact''), and our enhanced teacher (through knowledge aggregation, denoted by ``MT'') for reference. It should be noted that the results for DUKD, CrossKD, and three original pre-trained teacher/student models are from their corresponding publications, while the results for KD, AT and FAKD were generated by ourselves based on their publicly available source code. We did not report the results of CrossKD for the EDSR and SwinIR student model, and those of DUKD for EDSR student,  due to the unavailability of the source code and these results associated with their original papers. 

It can be observed from Table \ref{tbl:RCANResults}-\ref{tbl:SwinIRResults} that our MTKD approach consistently offers the best performance for all three student models, on four different test datasets, and for different scale factors. MTKD optimized RCAN\_lightweight model even outperforms its original full version (e.g., with a 0.26dB PSNR gain for scale $\times$4 on the Urban100 database). We also note that the performance improvement over other knowledge distillation methods is more significant when the performance difference between the original teacher and (pre-trained w/o KD) student models is larger - this aligns with the observation reported elsewhere \cite{miles2023understanding_AAAI}. The second best performer is DUKD for RCAN and FAKD for SwinIR and EDSR, typically with lower PSNR results (up to 0.46 dB) compared to MTKD. We also observe that most of the benchmarked knowledge distillation methods do offer improved performance compared to the corresponding pre-trained student model (without KD) - thus verifying the effectiveness of the knowledge distillation technique. As we mentioned above, the output of the Knowledge Aggregation module, $I^\mathrm{MT}_\mathrm{HR}$, has also been compared with its ground-truth counterpart, with results (denoted by ``MT'') are shown in  Table \ref{tbl:RCANResults}-\ref{tbl:SwinIRResults}. It is noted that for all three scale factors, and four test datasets, the MT results are always better than those for full EDSR, SwinIR and RCAN models, which also showcase the effectiveness of the Knowledge Aggregation module.

\subsection{Qualitative Evaluation}
\label{subsec:QualitativeE}

\begin{figure}[ht]
  \centering
  
    \rotatebox{90}{\small (a)}
        \begin{minipage}{0.2016\linewidth}
		  \centering
            \setlength{\abovecaptionskip}{0.cm}
		  \includegraphics[width=1\linewidth]{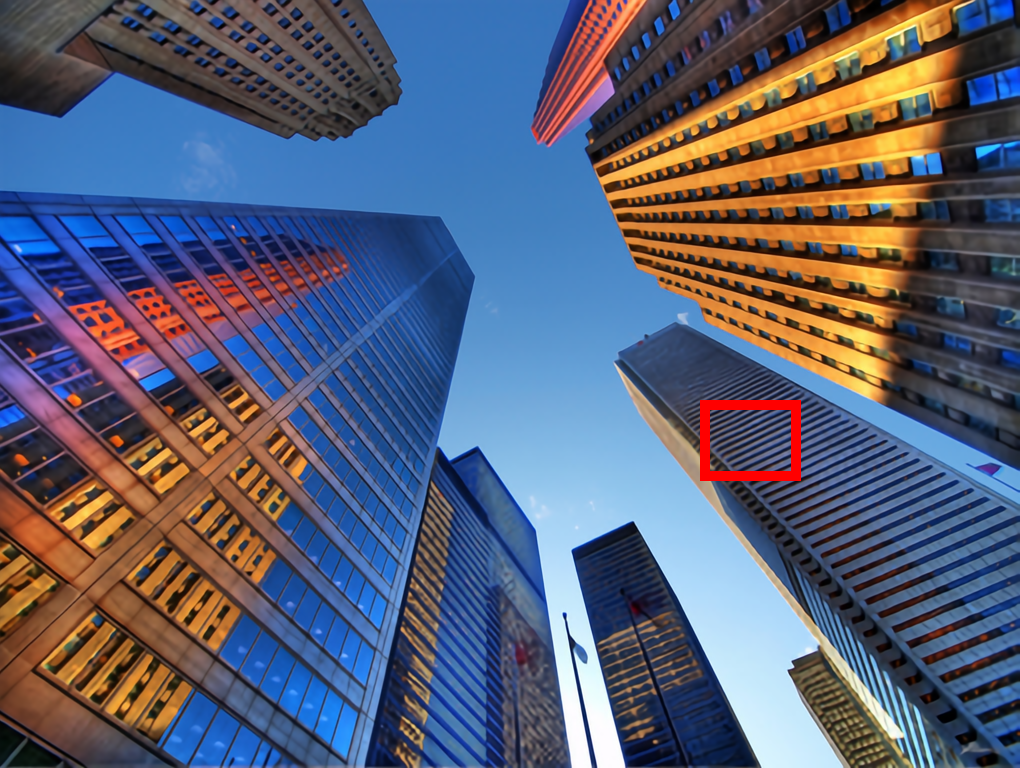}
            \caption*{\small img\_012 \protect\\ Urban100}
	  \end{minipage}
		\begin{minipage}{0.19\linewidth}
		  \centering
            \setlength{\abovecaptionskip}{0.cm}
		  \includegraphics[width=1\linewidth]{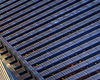}
            \caption*{\small HR \protect\\ PSNR/SSIM}
	  \end{minipage}
	  \begin{minipage}{0.19\linewidth}
		  \centering
            \setlength{\abovecaptionskip}{0.cm}
		  \includegraphics[width=1\linewidth]{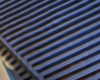}
            \caption*{\small Full \protect\\ 24.81/.7928}
	  \end{minipage}
 	\begin{minipage}{0.19\linewidth}
		  \centering
            \setlength{\abovecaptionskip}{0.cm}
		  \includegraphics[width=1\linewidth]{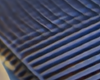}
            \caption*{\small Compact \protect\\ 24.20/.7542}
	  \end{minipage}

 	\begin{minipage}{0.18\linewidth}
		  \centering
            \setlength{\abovecaptionskip}{0.cm}
		  \includegraphics[width=1\linewidth]{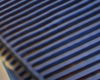}
            \caption*{\small MT \protect\\ 24.93/.7865}
	  \end{minipage}
        \begin{minipage}{0.18\linewidth}
		  \centering
            \setlength{\abovecaptionskip}{0.cm}
	      \includegraphics[width=1\linewidth]{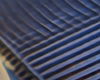}
            \caption*{\small KD \protect\\ 24.19/.7533 }
	  \end{minipage}
  	\begin{minipage}{0.18\linewidth}
		  \centering
            \setlength{\abovecaptionskip}{0.cm}
	      \includegraphics[width=1\linewidth]{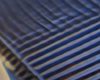}
            \caption*{\small AT \protect\\ 24.19/.7530 }
	  \end{minipage}
        \begin{minipage}{0.18\linewidth}
		  \centering
            \setlength{\abovecaptionskip}{0.cm}
	      \includegraphics[width=1\linewidth]{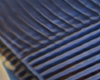}
            \caption*{\small FAKD \protect\\ 24.14/.7526}
	  \end{minipage}
        \begin{minipage}{0.18\linewidth}
		  \centering
            \setlength{\abovecaptionskip}{0.cm}
	      \includegraphics[width=1\linewidth]{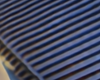}
            \caption*{\small Ours \protect\\ 24.34/.7622}
	  \end{minipage}

    \rotatebox{90}{\small (b)}
        \begin{minipage}{0.2286\linewidth}
		  \centering
            \setlength{\abovecaptionskip}{0.cm}
		  \includegraphics[width=1\linewidth]{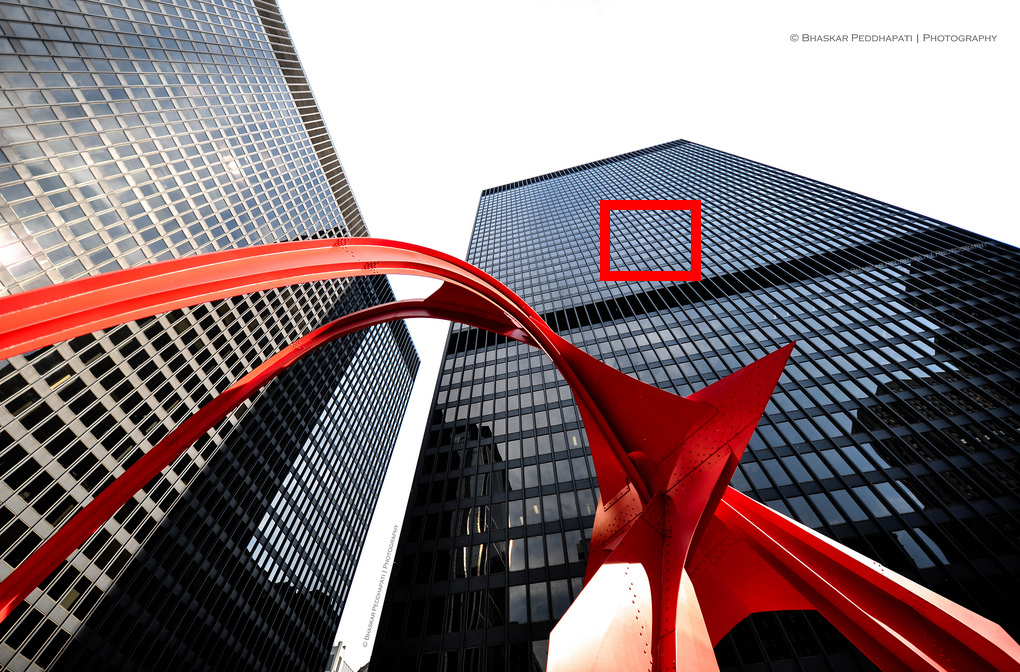}
            \caption*{\small img\_062 \protect\\ Urban100}
	  \end{minipage}
		\begin{minipage}{0.19\linewidth}
		  \centering
            \setlength{\abovecaptionskip}{0.cm}
		  \includegraphics[width=1\linewidth]{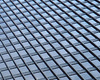}
            \caption*{\small HR \protect\\ PSNR/SSIM}
	  \end{minipage}
	  \begin{minipage}{0.19\linewidth}
		  \centering
            \setlength{\abovecaptionskip}{0.cm}
		  \includegraphics[width=1\linewidth]{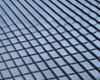}
            \caption*{\small Full \protect\\ 21.91/.9228}
	  \end{minipage}
 	\begin{minipage}{0.19\linewidth}
		  \centering
            \setlength{\abovecaptionskip}{0.cm}
		  \includegraphics[width=1\linewidth]{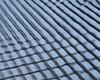}
            \caption*{\small Compact \protect\\ 20.48/.8920}
	  \end{minipage}

 	\begin{minipage}{0.18\linewidth}
		  \centering
            \setlength{\abovecaptionskip}{0.cm}
		  \includegraphics[width=1\linewidth]{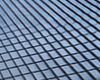}
            \caption*{\small MT \protect\\ 22.29/.9133}
	  \end{minipage}
        \begin{minipage}{0.18\linewidth}
		  \centering
            \setlength{\abovecaptionskip}{0.cm}
	      \includegraphics[width=1\linewidth]{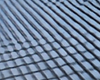}
            \caption*{\small KD \protect\\ 20.63/.8930}
	  \end{minipage}
  	\begin{minipage}{0.18\linewidth}
		  \centering
            \setlength{\abovecaptionskip}{0.cm}
	      \includegraphics[width=1\linewidth]{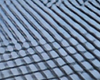}
            \caption*{\small AT \protect\\ 20.52/.8931}
	  \end{minipage}
        \begin{minipage}{0.18\linewidth}
		  \centering
            \setlength{\abovecaptionskip}{0.cm}
	      \includegraphics[width=1\linewidth]{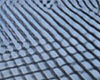}
            \caption*{\small FAKD \protect\\ 20.47/.8897}
	  \end{minipage}
        \begin{minipage}{0.18\linewidth}
		  \centering
            \setlength{\abovecaptionskip}{0.cm}
	      \includegraphics[width=1\linewidth]{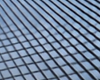}
            \caption*{\small Ours \protect\\ 21.43/.9064}
	  \end{minipage}

  \caption{The ×4 super-resolution results of \textbf{SwinIR} models on (a) img012, (b) img062 from Urban100. PSNRs and SSIMs are displayed below each image.}
  \label{fig:QualitativeESwinIRUrban100}
\end{figure}

\begin{figure}[!ht]
  \centering
    \rotatebox{90}{\small (a)}
        \begin{minipage}{0.2297\linewidth}
		  \centering
            \setlength{\abovecaptionskip}{0.cm}
		  \includegraphics[width=1\linewidth]{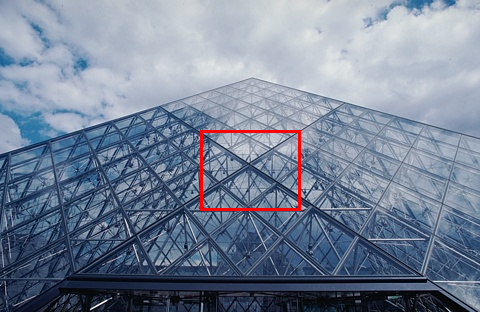}
            \caption*{\small img223061 \protect\\ BSD100}
	  \end{minipage}
		\begin{minipage}{0.19\linewidth}
		  \centering
            \setlength{\abovecaptionskip}{0.cm}
		  \includegraphics[width=1\linewidth]{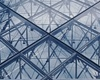}
            \caption*{\small HR \protect\\ PSNR/SSIM}
	  \end{minipage}
	  \begin{minipage}{0.19\linewidth}
		  \centering
            \setlength{\abovecaptionskip}{0.cm}
		  \includegraphics[width=1\linewidth]{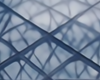}
            \caption*{\small Full \protect\\ 25.34/.7300}
	  \end{minipage}
 	\begin{minipage}{0.19\linewidth}
		  \centering
            \setlength{\abovecaptionskip}{0.cm}
		  \includegraphics[width=1\linewidth]{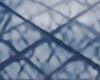}
            \caption*{\small Compact \protect\\ 24.86/.6932}
	  \end{minipage}
   
 	\begin{minipage}{0.18\linewidth}
		  \centering
            \setlength{\abovecaptionskip}{0.cm}
		  \includegraphics[width=1\linewidth]{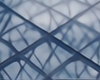}
            \caption*{\small MT \protect\\ 25.38/.7305}
	  \end{minipage}
        \begin{minipage}{0.18\linewidth}
		  \centering
            \setlength{\abovecaptionskip}{0.cm}
	      \includegraphics[width=1\linewidth]{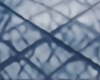}
            \caption*{\small KD \protect\\ 24.87/.6925}
	  \end{minipage}
  	\begin{minipage}{0.18\linewidth}
		  \centering
            \setlength{\abovecaptionskip}{0.cm}
	      \includegraphics[width=1\linewidth]{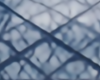}
            \caption*{\small AT \protect\\ 24.85/.6903}
	  \end{minipage}
        \begin{minipage}{0.18\linewidth}
		  \centering
            \setlength{\abovecaptionskip}{0.cm}
	      \includegraphics[width=1\linewidth]{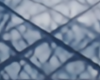}
            \caption*{\small FAKD \protect\\ 24.85/.6902}
	  \end{minipage}
        \begin{minipage}{0.18\linewidth}
		  \centering
            \setlength{\abovecaptionskip}{0.cm}
	      \includegraphics[width=1\linewidth]{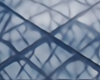}
            \caption*{\small Ours \protect\\ 25.00/.7050}
	  \end{minipage}

    \rotatebox{90}{\small (b)}
        \begin{minipage}{0.2297\linewidth}
		  \centering
            \setlength{\abovecaptionskip}{0.cm}
		  \includegraphics[width=1\linewidth]{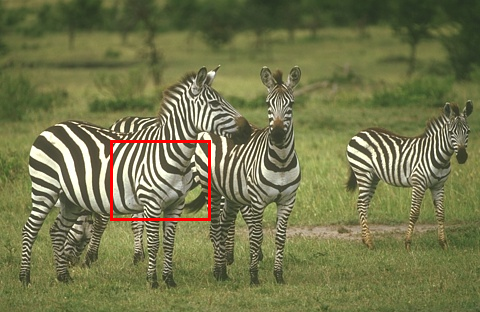}
            \caption*{\small img253027 \protect\\ BSD100}
	  \end{minipage}
		\begin{minipage}{0.19\linewidth}
		  \centering
            \setlength{\abovecaptionskip}{0.cm}
		  \includegraphics[width=1\linewidth]{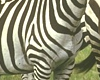}
            \caption*{\small HR \protect\\ PSNR/SSIM}
	  \end{minipage}
	  \begin{minipage}{0.19\linewidth}
		  \centering
            \setlength{\abovecaptionskip}{0.cm}
		  \includegraphics[width=1\linewidth]{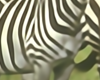}
            \caption*{\small Full \protect\\ 23.06/.7257}
	  \end{minipage}
 	\begin{minipage}{0.19\linewidth}
		  \centering
            \setlength{\abovecaptionskip}{0.cm}
		  \includegraphics[width=1\linewidth]{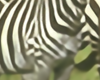}
            \caption*{\small Compact \protect\\ 22.81/.7134}
	  \end{minipage}

 	\begin{minipage}{0.18\linewidth}
		  \centering
            \setlength{\abovecaptionskip}{0.cm}
		  \includegraphics[width=1\linewidth]{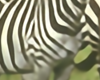}
            \caption*{\small MT \protect\\ 23.16/.7254}
	  \end{minipage}
        \begin{minipage}{0.18\linewidth}
		  \centering
            \setlength{\abovecaptionskip}{0.cm}
	      \includegraphics[width=1\linewidth]{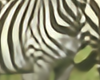}
            \caption*{\small KD \protect\\ 22.84/.7139}
	  \end{minipage}
  	\begin{minipage}{0.18\linewidth}
		  \centering
            \setlength{\abovecaptionskip}{0.cm}
	      \includegraphics[width=1\linewidth]{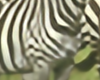}
            \caption*{\small AT \protect\\ 22.84/.7136}
	  \end{minipage}
        \begin{minipage}{0.18\linewidth}
		  \centering
            \setlength{\abovecaptionskip}{0.cm}
	      \includegraphics[width=1\linewidth]{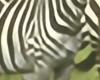}
            \caption*{\small FAKD \protect\\ 22.78/.7127}
	  \end{minipage}
        \begin{minipage}{0.18\linewidth}
		  \centering
            \setlength{\abovecaptionskip}{0.cm}
	      \includegraphics[width=1\linewidth]{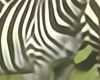}
            \caption*{\small Ours \protect\\ 22.98/.7199}
	  \end{minipage}

  \caption{The ×4 super-resolution results of \textbf{SwinIR} models on (a) img223061 and (b) img253027 from BSD100. PSNRs and SSIMs are displayed below each image.}
  \label{fig:QualitativeESwinIRBSD100}
\end{figure}

Fig. \ref{fig:QualitativeESwinIRUrban100} and \ref{fig:QualitativeESwinIRBSD100} shows a visual comparison between results generated by various SwinIR models (for a scale factor 4) that are trained using different knowledge distillation methods. The source images presented are from the Urban100 dataset, which is the most challenging one among all four test sets. It can be observed that for repetitive texture reconstruction (subfigure (a) and (b)) in Fig. \ref{fig:QualitativeESwinIRUrban100}, MTKD produces sharper edges and structures which are closer to the ground-truth high-resolution images compared to other knowledge distillation methods. This may be due to the use of the DWT-based loss function, which is able to perceive high frequency energy changes at various directions. MTKD can also generate more spatial details as shown in subfigures (c) and (d) in Fig. \ref{fig:QualitativeESwinIRBSD100}, such as complex textures. The perceptual quality enhancement achieved by MTKD has also been verified by the corresponding PSNR values in each example.

\subsection{Ablation Study}
\label{sec:AblationStudy}

To demonstrate the effectiveness of each primary contribution in this work, we conducted ablation studies to more fully characterise the impact of our proposed MTKD framework.

\subsubsection{Study 1: Multiple teachers}
\label{subsec:Study1}

To confirm the contribution of multiple teachers in the MTKD framework, we created two different variants, each of which employs 1-2 teachers. Here only the RCAN\_lightweight model (for the $\times 4$ ISR task) is employed in this experiment, and the created variants include (v1) with both SwinIR and EDSR as teachers; (v2) only with SwinIR as teacher. Here we kept SwinIR as a teacher in both variants due to its superior performance over the other two teachers. As shown in Table \ref{tbl:study}, the performance is improved in line with the number of teachers, with the full MTKD method (with three teachers) delivering the best ISR results. 

\begin{figure*}[ht]
	\centering

		\rotatebox{90}{\small (a)}
		\begin{minipage}{0.17\linewidth}
		  \centering
            \setlength{\abovecaptionskip}{0.cm}
		  \includegraphics[width=1\linewidth]{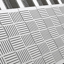}
            \caption*{\small LR \protect\\ PSNR/SSIM}
	  \end{minipage}
	  \begin{minipage}{0.17\linewidth}
		  \centering
            \setlength{\abovecaptionskip}{0.cm}
		  \includegraphics[width=1\linewidth]{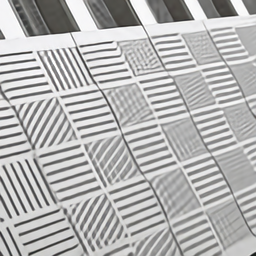}
            \caption*{\small EDSR \protect\\ 14.13/0.44}
	  \end{minipage}
 	\begin{minipage}{0.17\linewidth}
		  \centering
            \setlength{\abovecaptionskip}{0.cm}
		  \includegraphics[width=1\linewidth]{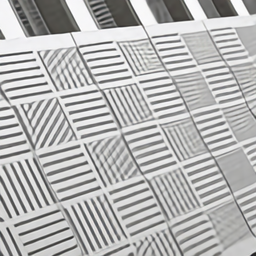}
            \caption*{\small RCAN \protect\\ 14.55/0.50}
	  \end{minipage}
 	\begin{minipage}{0.17\linewidth}
		  \centering
            \setlength{\abovecaptionskip}{0.cm}
		  \includegraphics[width=1\linewidth]{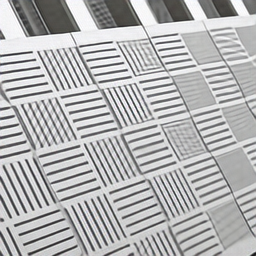}
            \caption*{\small SwinIR \protect\\ 16.06/0.64}
	  \end{minipage}
  	\begin{minipage}{0.17\linewidth}
		  \centering
            \setlength{\abovecaptionskip}{0.cm}
	      \includegraphics[width=1\linewidth]{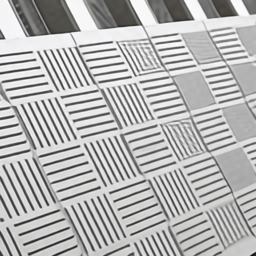}
            \caption*{\small MT \protect\\ 16.52/0.68 }
	  \end{minipage}

		\rotatebox{90}{\small{(b)}}
		\begin{minipage}{0.17\linewidth}
		  \centering
            \setlength{\abovecaptionskip}{0.cm}
		  \includegraphics[width=1\linewidth]{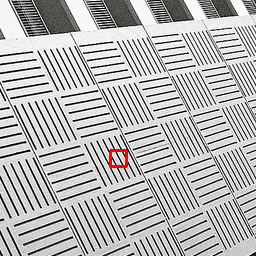}
            \caption*{\small HR \protect\\ (DI)}
	  \end{minipage}
        \begin{minipage}{0.17\linewidth}
		  \centering
            \setlength{\abovecaptionskip}{0.cm}
		  \includegraphics[width=1\linewidth]{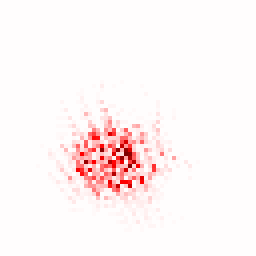}
            \caption*{\small EDSR \protect\\ (6.75)}
	  \end{minipage}
	  \begin{minipage}{0.17\linewidth}
		  \centering
            \setlength{\abovecaptionskip}{0.cm}
		  \includegraphics[width=1\linewidth]{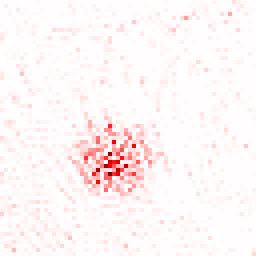}
            \caption*{\small RCAN \protect\\ (24.42)}
	  \end{minipage}
        \begin{minipage}{0.17\linewidth}
		  \centering
            \setlength{\abovecaptionskip}{0.cm}
		  \includegraphics[width=1\linewidth]{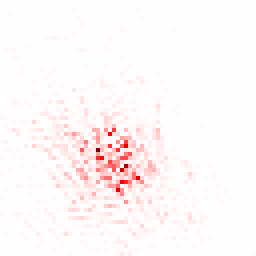}
            \caption*{\small SwinIR \protect\\ (16.66)}
	  \end{minipage}
        \begin{minipage}{0.17\linewidth}
		  \centering
            \setlength{\abovecaptionskip}{0.cm}
		  \includegraphics[width=1\linewidth]{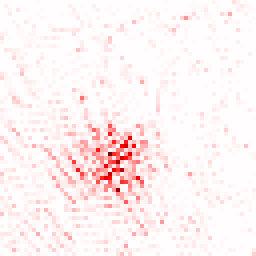}
            \caption*{\small MT\protect\\ (26.09) }
	  \end{minipage}


    \caption{The illustration of the contribution from each teacher model using the Local Attribution Maps tool \cite{gu2021interpreting}. (a) The input low-resolution image and the high resolution reconstructions generated by the original EDSR, RCAN, and SwinIR at a $\times4$ scale. (b) the high-resolution image and LAM \cite{gu2021interpreting} maps for all three teacher models. Here diffusion Index (DI) quantifies the overall contribution from each teacher.}
    \label{Fig:LAM}
\end{figure*}

To further showcase the contribution of each employed teacher model in the proposed framework, we utilize the Local Attribution Maps tool \cite{gu2021interpreting} to identify the level of contribution from the pixels in each teacher's output, $I^n_\mathrm{HR}$, to those in the output of the Knowledge Aggregation module, $I^\mathrm{MT}_\mathrm{HR}$. As illustrated in Fig. \ref{Fig:LAM}, we can observe that all three teacher models have contributed information to the final output of the Knowledge Aggregation module.

\begin{table}[!ht]
\centering
\setlength\tabcolsep{6pt}
\caption{Ablation study results with RCAN\_lightweight ($\times$4) as the student model.}
\begin{tabular}{r|ccccc}
\toprule
Variants &  v1 & v2 & v3 & v4 \\ \midrule
PSNR/SSIM & 26.79/0.8018 & 26.70/0.8003 & 26.93/0.8062 & 26.89/0.8043 \\
\midrule \midrule
Variants & v5 & v6 & v7 & \textbf{Ours}\\ \midrule
PSNR/SSIM  & 26.78/0.8023 & 26.98/0.8078 & 26.28/0.7958 & 27.08/0.8108\\ \bottomrule
\end{tabular}
\label{tbl:study}
\end{table}

\subsubsection{Study 2 - Knowledge aggregation network structure}
\label{subsec:Study2}
As the design of the Knowledge Aggregation network (with the new DCTSwin blocks) is one of the primary contributions in this work, we tested the effectiveness of the proposed DCTSwin blocks by replacing them with Mixer Layer blocks \cite{tolstikhin2021mlp} (v3) and removing the DCT and IDCT modules (v4). In (v3), to enable a fair comparison, we keep the number of parameters similar to that of the original knowledge aggregation network. The experiment is also based on the Urban100 database, RCAN\_lightweight model and the scale factor 4. Based on the results summarized in Table \ref{tbl:study}, both (v3) and (v4) have been outperformed by the original network design, which shows the importance of using DCT/IDCT modules and the DCTSwin blocks.

\subsubsection{Study 3 - Distillation loss function}
\label{subsec:Study3}

To validate the effectiveness of the wavelet-based distillation loss, we have replaced it with L1 loss (v5), DCT-based loss \cite{tomosada2021gan} (v6) and an alternative DWT loss \cite{zhang2022wavelet} (which only focuses on high-frequency information) (v7) but kept the network architectures and training configurations the same. This study is based on the Urban100 dataset and the RCAN\_lightweight model for the $\times 4$ task ISR task. The results shown in Table \ref{tbl:study} indicate that the employed wavelet-based distillation loss does improve the model performance compared to other tested loss functions.

\section{Conclusion}
\label{sec:Conclusion}

This paper presents a novel Multi-Teacher Knowledge Distillation (MTKD) framework for image super-resolution. The proposed approach integrates a new DCTSwin-based network to aggregate the knowledge from multiple teacher models and generates an enhanced representation of the high-resolution image. This is employed to optimize the student model through distillation using a loss function based on discrete wavelet transform. We conduct comprehensive experiments for the image super-resolution task using various teacher and student networks and diverse test databases. Throughout these experiments, our method consistently outperforms other existing knowledge distillation methods, showcasing its effectiveness and robustness for the ISR task. The primary contributions of this work have also been verified in the additional ablation study. Future work should focus on the application of this approach to other low-level computer vision tasks.

\section{Acknowledgements}
The authors appreciate the funding from Netflix Inc., University of Bristol, and the UKRI MyWorld Strength in Places Programme (SIPF00006/1).

\clearpage  

%
%
\bibliographystyle{splncs04}
\small
\setlength{\bibsep}{0.5em} 
\bibliography{egbib}

\begin{thebibliography}{10}
\providecommand{\url}[1]{\texttt{#1}}
\providecommand{\urlprefix}{URL }
\providecommand{\doi}[1]{https://doi.org/#1}

\bibitem{afonso2018video}
Afonso, M., Zhang, F., Bull, D.R.: Video compression based on spatio-temporal resolution adaptation. IEEE Transactions on Circuits and Systems for Video Technology  \textbf{29}(1),  275--280 (2018)

\bibitem{aleissaee2023transformers}
Aleissaee, A.A., Kumar, A., Anwer, R.M., Khan, S., Cholakkal, H., Xia, G.S., Khan, F.S.: Transformers in remote sensing: A survey. Remote Sensing  \textbf{15}(7), ~1860 (2023)

\bibitem{bevilacqua2012low}
Bevilacqua, M., Roumy, A., Guillemot, C., Morel, M.L.A.: Low-complexity single-image super-resolution based on nonnegative neighbor embedding. In: British Machine Vision Conference (BMVC) (2012)

\bibitem{buciluǎ2006model}
Buciluǎ, C., Caruana, R., Niculescu-Mizil, A.: Model compression. In: Proceedings of the 12th ACM SIGKDD international conference on Knowledge discovery and data mining. pp. 535--541 (2006)

\bibitem{bull2021intelligent}
Bull, D., Zhang, F.: Intelligent image and video compression: communicating pictures. Academic Press (2021)

\bibitem{cai2022real}
Cai, J., Meng, Z., Ding, J., Ho, C.M.: Real-time super-resolution for real-world images on mobile devices. In: 2022 IEEE 5th International Conference on Multimedia Information Processing and Retrieval (MIPR). pp. 127--132. IEEE (2022)

\bibitem{chawla2021data}
Chawla, A., Yin, H., Molchanov, P., Alvarez, J.: Data-free knowledge distillation for object detection. In: Proceedings of the IEEE/CVF Winter Conference on Applications of Computer Vision. pp. 3289--3298 (2021)

\bibitem{chebotar2016distilling}
Chebotar, Y., Waters, A.: Distilling knowledge from ensembles of neural networks for speech recognition. In: Interspeech. pp. 3439--3443 (2016)

\bibitem{chen2017learning}
Chen, G., Choi, W., Yu, X., Han, T., Chandraker, M.: Learning efficient object detection models with knowledge distillation. Advances in neural information processing systems  \textbf{30} (2017)

\bibitem{chen2021distilling}
Chen, P., Liu, S., Zhao, H., Jia, J.: Distilling knowledge via knowledge review. In: Proceedings of the IEEE/CVF Conference on Computer Vision and Pattern Recognition. pp. 5008--5017 (2021)

\bibitem{chen2023hat}
Chen, X., Wang, X., Zhang, W., Kong, X., Qiao, Y., Zhou, J., Dong, C.: Hat: Hybrid attention transformer for image restoration. arXiv preprint arXiv:2309.05239  (2023)

\bibitem{chira2022image}
Chira, D., Haralampiev, I., Winther, O., Dittadi, A., Li{\'e}vin, V.: Image super-resolution with deep variational autoencoders. In: European Conference on Computer Vision. pp. 395--411. Springer (2022)

\bibitem{conde2022swin2sr}
Conde, M.V., Choi, U.J., Burchi, M., Timofte, R.: {Swin2SR}: Swinv2 transformer for compressed image super-resolution and restoration. In: European Conference on Computer Vision. pp. 669--687. Springer (2022)

\bibitem{dong2015image}
Dong, C., Loy, C.C., He, K., Tang, X.: Image super-resolution using deep convolutional networks. IEEE transactions on pattern analysis and machine intelligence  \textbf{38}(2),  295--307 (2015)

\bibitem{fang2022cross}
Fang, H., Hu, X., Hu, H.: Cross knowledge distillation for image super-resolution. In: Proceedings of the 2022 6th International Conference on Video and Image Processing. pp. 162--168 (2022)

\bibitem{feng2023rankdvqa}
Feng, C., Danier, D., Wang, H., Zhang, F., Bull, D.: {RankDVQA-mini}: Knowledge distillation-driven deep video quality assessment. arXiv preprint arXiv:2312.08864  (2023)

\bibitem{gao2018image}
Gao, Q., Zhao, Y., Li, G., Tong, T.: Image super-resolution using knowledge distillation. In: Asian Conference on Computer Vision. pp. 527--541. Springer (2018)

\bibitem{gao2023implicit}
Gao, S., Liu, X., Zeng, B., Xu, S., Li, Y., Luo, X., Liu, J., Zhen, X., Zhang, B.: Implicit diffusion models for continuous super-resolution. In: Proceedings of the IEEE/CVF Conference on Computer Vision and Pattern Recognition. pp. 10021--10030 (2023)

\bibitem{gu2021interpreting}
Gu, J., Dong, C.: Interpreting super-resolution networks with local attribution maps. In: Proceedings of the IEEE/CVF Conference on Computer Vision and Pattern Recognition. pp. 9199--9208 (2021)

\bibitem{gu2021class}
Gu, Y., Deng, C., Wei, K.: Class-incremental instance segmentation via multi-teacher networks. In: Proceedings of the AAAI Conference on Artificial Intelligence. vol.~35, pp. 1478--1486 (2021)

\bibitem{he2020fakd}
He, Z., Dai, T., Lu, J., Jiang, Y., Xia, S.T.: {FAKD}: Feature-affinity based knowledge distillation for efficient image super-resolution. In: 2020 IEEE International Conference on Image Processing (ICIP). pp. 518--522. IEEE (2020)

\bibitem{heil1989continuous}
Heil, C.E., Walnut, D.F.: Continuous and discrete wavelet transforms. SIAM review  \textbf{31}(4),  628--666 (1989)

\bibitem{hinton2015distilling}
Hinton, G., Vinyals, O., Dean, J.: Distilling the knowledge in a neural network. arXiv preprint arXiv:1503.02531  (2015)

\bibitem{huang2015single}
Huang, J.B., Singh, A., Ahuja, N.: Single image super-resolution from transformed self-exemplars. In: Proceedings of the IEEE conference on computer vision and pattern recognition. pp. 5197--5206 (2015)

\bibitem{jacob2023online}
Jacob, G.M., Agarwal, V., Stenger, B.: Online knowledge distillation for multi-task learning. In: Proceedings of the IEEE/CVF Winter Conference on Applications of Computer Vision. pp. 2359--2368 (2023)

\bibitem{jeevan2024wavemixsr}
Jeevan, P., Srinidhi, A., Prathiba, P., Sethi, A.: {WaveMixSR}: Resource-efficient neural network for image super-resolution. In: Proceedings of the IEEE/CVF Winter Conference on Applications of Computer Vision. pp. 5884--5892 (2024)

\bibitem{jeevan2022wavemix}
Jeevan, P., Viswanathan, K., Sethi, A.: {WaveMix}: A resource-efficient neural network for image analysis. arXiv preprint arXiv:2205.14375  (2022)

\bibitem{jiang2023compressing}
Jiang, Y., Nawala, J., Zhang, F., Bull, D.: Compressing deep image super-resolution models. arXiv preprint arXiv:2401.00523  (2023)

\bibitem{jiao2019tinybert}
Jiao, X., Yin, Y., Shang, L., Jiang, X., Chen, X., Li, L., Wang, F., Liu, Q.: Tinybert: Distilling bert for natural language understanding. arXiv preprint arXiv:1909.10351  (2019)

\bibitem{jin2023multi}
Jin, Y., Wang, J., Lin, D.: Multi-level logit distillation. In: Proceedings of the IEEE/CVF Conference on Computer Vision and Pattern Recognition. pp. 24276--24285 (2023)

\bibitem{kang2023distill}
Kang, J., Xu, W., Ritter, A.: Distill or annotate? {Cost-efficient} fine-tuning of compact models. arXiv preprint arXiv:2305.01645  (2023)

\bibitem{kim2016accurate}
Kim, J., Lee, J.K., Lee, K.M.: Accurate image super-resolution using very deep convolutional networks. In: Proceedings of the IEEE conference on computer vision and pattern recognition. pp. 1646--1654 (2016)

\bibitem{kingma2014adam}
Kingma, D.P., Ba, J.: Adam: a method for stochastic optimization. arXiv preprint arXiv:1412.6980  (2014)

\bibitem{ledig2017photo}
Ledig, C., Theis, L., Husz{\'a}r, F., Caballero, J., Cunningham, A., Acosta, A., Aitken, A., Tejani, A., Totz, J., Wang, Z., et~al.: Photo-realistic single image super-resolution using a generative adversarial network. In: Proceedings of the IEEE conference on computer vision and pattern recognition. pp. 4681--4690 (2017)

\bibitem{lee2020learning}
Lee, W., Lee, J., Kim, D., Ham, B.: Learning with privileged information for efficient image super-resolution. In: Computer Vision--ECCV 2020: 16th European Conference, Glasgow, UK, August 23--28, 2020, Proceedings, Part XXIV 16. pp. 465--482. Springer (2020)

\bibitem{lee2105fnet}
Lee-Thorp, J., Ainslie, J., Eckstein, I., Ontanon, S.: {Fnet}: Mixing tokens with fourier transforms. arXiv preprint arXiv:2105.03824  (2021)

\bibitem{lepcha2023image}
Lepcha, D.C., Goyal, B., Dogra, A., Goyal, V.: Image super-resolution: A comprehensive review, recent trends, challenges and applications. Information Fusion  \textbf{91},  230--260 (2023)

\bibitem{li2020knowledge}
Li, W.H., Bilen, H.: Knowledge distillation for multi-task learning. In: Computer Vision--ECCV 2020 Workshops: Glasgow, UK, August 23--28, 2020, Proceedings, Part VI 16. pp. 163--176. Springer (2020)

\bibitem{liang2021swinir}
Liang, J., Cao, J., Sun, G., Zhang, K., Van~Gool, L., Timofte, R.: {SwinIR}: Image restoration using swin transformer. In: Proceedings of the IEEE/CVF international conference on computer vision. pp. 1833--1844 (2021)

\bibitem{lim2017enhanced}
Lim, B., Son, S., Kim, H., Nah, S., Mu~Lee, K.: Enhanced deep residual networks for single image super-resolution. In: Proceedings of the IEEE conference on computer vision and pattern recognition workshops. pp. 136--144 (2017)

\bibitem{lin2022knowledge}
Lin, S., Xie, H., Wang, B., Yu, K., Chang, X., Liang, X., Wang, G.: Knowledge distillation via the target-aware transformer. In: Proceedings of the IEEE/CVF Conference on Computer Vision and Pattern Recognition. pp. 10915--10924 (2022)

\bibitem{lin2022revisiting}
Lin, Z., Garg, P., Banerjee, A., Magid, S.A., Sun, D., Zhang, Y., Van~Gool, L., Wei, D., Pfister, H.: Revisiting rcan: Improved training for image super-resolution. arXiv preprint arXiv:2201.11279  (2022)

\bibitem{liu2023pre}
Liu, P., Yuan, W., Fu, J., Jiang, Z., Hayashi, H., Neubig, G.: Pre-train, prompt, and predict: A systematic survey of prompting methods in natural language processing. ACM Computing Surveys  \textbf{55}(9),  1--35 (2023)

\bibitem{liu2021swin}
Liu, Z., Lin, Y., Cao, Y., Hu, H., Wei, Y., Zhang, Z., Lin, S., Guo, B.: {Swin Transformer}: Hierarchical vision transformer using shifted windows. In: Proceedings of the IEEE/CVF international conference on computer vision. pp. 10012--10022 (2021)

\bibitem{liu2020photo}
Liu, Z.S., Siu, W.C., Chan, Y.L.: Photo-realistic image super-resolution via variational autoencoders. IEEE Transactions on Circuits and Systems for video Technology  \textbf{31}(4),  1351--1365 (2020)

\bibitem{lopez2022attention}
L{\'o}pez-Cifuentes, A., Escudero-Vi{\~n}olo, M., Besc{\'o}s, J., SanMiguel, J.C.: Attention-based knowledge distillation in multi-attention tasks: The impact of a {DCT}-driven loss. arXiv preprint arXiv:2205.01997  (2022)

\bibitem{lu2022transformer}
Lu, Z., Li, J., Liu, H., Huang, C., Zhang, L., Zeng, T.: Transformer for single image super-resolution. In: Proceedings of the IEEE/CVF conference on computer vision and pattern recognition. pp. 457--466 (2022)

\bibitem{martin2001database}
Martin, D., Fowlkes, C., Tal, D., Malik, J.: A database of human segmented natural images and its application to evaluating segmentation algorithms and measuring ecological statistics. In: Proceedings Eighth IEEE International Conference on Computer Vision. ICCV 2001. vol.~2, pp. 416--423. IEEE (2001)

\bibitem{meng2021multi}
Meng, Z., Yao, X., Sun, L.: Multi-task distillation: Towards mitigating the negative transfer in multi-task learning. In: 2021 IEEE International Conference on Image Processing (ICIP). pp. 389--393. IEEE (2021)

\bibitem{miles2023understanding_AAAI}
Miles, R., Mikolajczyk, K.: Understanding the role of the projector in knowledge distillation. In: Proceedings of the 38th AAAI Conference on Artificial Intelligence (AAAI-24) (December 2023)

\bibitem{morris2023st}
Morris, C., Danier, D., Zhang, F., Anantrasirichai, N., Bull, D.R.: {ST-MFNet Mini}: Knowledge distillation-driven frame interpolation. arXiv preprint arXiv:2302.08455  (2023)

\bibitem{park2023content}
Park, J., Son, S., Lee, K.M.: Content-aware local gan for photo-realistic super-resolution. In: Proceedings of the IEEE/CVF International Conference on Computer Vision. pp. 10585--10594 (2023)

\bibitem{peng2023accelerating}
Peng, T., Gao, G., Sun, H., Zhang, F., Bull, D.: Accelerating learnt video codecs with gradient decay and layer-wise distillation. arXiv preprint arXiv:2312.02605  (2023)

\bibitem{saharia2022image}
Saharia, C., Ho, J., Chan, W., Salimans, T., Fleet, D.J., Norouzi, M.: Image super-resolution via iterative refinement. IEEE Transactions on Pattern Analysis and Machine Intelligence  \textbf{45}(4),  4713--4726 (2022)

\bibitem{shi2020proxylesskd}
Shi, W., Ren, G., Chen, Y., Yan, S.: {ProxylessKD}: Direct knowledge distillation with inherited classifier for face recognition. arXiv preprint arXiv:2011.00265  (2020)

\bibitem{shi2016real}
Shi, W., Caballero, J., Husz{\'a}r, F., Totz, J., Aitken, A.P., Bishop, R., Rueckert, D., Wang, Z.: Real-time single image and video super-resolution using an efficient sub-pixel convolutional neural network. In: Proceedings of the IEEE conference on computer vision and pattern recognition. pp. 1874--1883 (2016)

\bibitem{singh2014various}
Singh, G., Mittal, A.: Various image enhancement techniques-a critical review. International Journal of Innovation and Scientific Research  \textbf{10}(2),  267--274 (2014)

\bibitem{timofte2017ntire}
Timofte, R., Agustsson, E., Van~Gool, L., Yang, M.H., Zhang, L.: {NTIRE} 2017 challenge on single image super-resolution: Methods and results. In: Proceedings of the IEEE conference on computer vision and pattern recognition workshops. pp. 114--125 (2017)

\bibitem{tolstikhin2021mlp}
Tolstikhin, I.O., Houlsby, N., Kolesnikov, A., Beyer, L., Zhai, X., Unterthiner, T., Yung, J., Steiner, A., Keysers, D., Uszkoreit, J., et~al.: Mlp-mixer: An all-mlp architecture for vision. Advances in neural information processing systems  \textbf{34},  24261--24272 (2021)

\bibitem{tomosada2021gan}
Tomosada, H., Kudo, T., Fujisawa, T., Ikehara, M.: Gan-based image deblurring using {DCT} loss with customized datasets. IEEE Access  \textbf{9},  135224--135233 (2021)

\bibitem{vaswani2017attention}
Vaswani, A., Shazeer, N., Parmar, N., Uszkoreit, J., Jones, L., Gomez, A.N., Kaiser, {\L}., Polosukhin, I.: Attention is all you need. Advances in neural information processing systems  \textbf{30} (2017)

\bibitem{basicsr}
Wang, X., Xie, L., Yu, K., Chan, K.C., Loy, C.C., Dong, C.: {BasicSR}: Open source image and video restoration toolbox. \url{https://github.com/XPixelGroup/BasicSR} (2022)

\bibitem{wang2020deep}
Wang, Z., Chen, J., Hoi, S.C.: Deep learning for image super-resolution: A survey. IEEE transactions on pattern analysis and machine intelligence  \textbf{43}(10),  3365--3387 (2020)

\bibitem{wang2004image}
Wang, Z., Bovik, A.C., Sheikh, H.R., Simoncelli, E.P.: {Image Quality Assessment}: from error visibility to structural similarity. IEEE transactions on image processing  \textbf{13}(4),  600--612 (2004)

\bibitem{wu2019multi}
Wu, M.C., Chiu, C.T., Wu, K.H.: Multi-teacher knowledge distillation for compressed video action recognition on deep neural networks. In: ICASSP 2019-2019 IEEE International Conference on Acoustics, Speech and Signal Processing (ICASSP). pp. 2202--2206. IEEE (2019)

\bibitem{xu2022dct}
Xu, R., Kang, X., Li, C., Chen, H., Ming, A.: {DCT-FANet}: {DCT} based frequency attention network for single image super-resolution. Displays  \textbf{74},  102220 (2022)

\bibitem{yadav2021frequency}
Yadav, O., Ghosal, K., Lutz, S., Smolic, A.: Frequency-domain loss function for deep exposure correction of dark images. Signal, Image and Video Processing  \textbf{15}(8),  1829--1836 (2021)

\bibitem{yang2017image}
Yang, J., Huang, T.: Image super-resolution: Historical overview and future challenges. In: Super-resolution imaging, pp. 1--34. CRC Press (2017)

\bibitem{yuan2021reinforced}
Yuan, F., Shou, L., Pei, J., Lin, W., Gong, M., Fu, Y., Jiang, D.: Reinforced multi-teacher selection for knowledge distillation. In: Proceedings of the AAAI Conference on Artificial Intelligence. vol.~35, pp. 14284--14291 (2021)

\bibitem{zagoruyko2016paying}
Zagoruyko, S., Komodakis, N.: Paying more attention to attention: Improving the performance of convolutional neural networks via attention transfer. arXiv preprint arXiv:1612.03928  (2016)

\bibitem{zeyde2012single}
Zeyde, R., Elad, M., Protter, M.: On single image scale-up using sparse-representations. In: Curves and Surfaces: 7th International Conference, Avignon, France, June 24-30, 2010, Revised Selected Papers 7. pp. 711--730. Springer (2012)

\bibitem{zhang2022swinfir}
Zhang, D., Huang, F., Liu, S., Wang, X., Jin, Z.: {SwinFIR}: Revisiting the swinir with fast {Fourier} convolution and improved training for image super-resolution. arXiv preprint arXiv:2208.11247  (2022)

\bibitem{zhang2022wavelet}
Zhang, L., Chen, X., Tu, X., Wan, P., Xu, N., Ma, K.: Wavelet knowledge distillation: Towards efficient image-to-image translation. In: Proceedings of the IEEE/CVF Conference on Computer Vision and Pattern Recognition. pp. 12464--12474 (2022)

\bibitem{zhang2018image}
Zhang, Y., Li, K., Li, K., Wang, L., Zhong, B., Fu, Y.: Image super-resolution using very deep residual channel attention networks. In: Proceedings of the European conference on computer vision (ECCV). pp. 286--301 (2018)

\bibitem{zhang2023data}
Zhang, Y., Li, W., Li, S., Hu, J., Chen, H., Wang, H., Tu, Z., Wang, W., Jing, B., Wang, Y.: Data upcycling knowledge distillation for image super-resolution. arXiv preprint arXiv:2309.14162  (2023)

\bibitem{zhu2018knowledge}
Zhu, X., Gong, S., et~al.: Knowledge distillation by on-the-fly native ensemble. Advances in neural information processing systems  \textbf{31} (2018)

\bibitem{zhu2023scalekd}
Zhu, Y., Zhou, Q., Liu, N., Xu, Z., Ou, Z., Mou, X., Tang, J.: {ScaleKD}: Distilling scale-aware knowledge in small object detector. In: Proceedings of the IEEE/CVF Conference on Computer Vision and Pattern Recognition. pp. 19723--19733 (2023)

\end{thebibliography}
\end{document}